\documentclass[prd, nofootinbib]{revtex4}
\usepackage{graphicx}
\usepackage{dcolumn}
\usepackage{slashbox,multirow}
\usepackage{amsmath,amsthm,amssymb}

\begin{document}
%


\title{A note about late-time wave tails on a dynamical background }

\author{Piotr Bizo\'n}
\affiliation{M. Smoluchowski Institute of Physics, Jagiellonian
University, Krak\'ow, Poland}
\author{Andrzej Rostworowski}
\affiliation{M. Smoluchowski Institute of Physics, Jagiellonian
University, Krak\'ow, Poland}

\date{\today}
\begin{abstract}
Consider a spherically symmetric spacetime generated by a self-gravitating massless scalar field
$\phi$ and let $\psi$ be a test (nonspherical) massless scalar field propagating on this
dynamical background. Gundlach, Price, and Pullin \cite{gpp2} computed numerically the late-time
tails for different multipoles of the  field $\psi$ and suggested that solutions with compactly
supported initial data  decay in accord with Price's law as $t^{-(2\ell+3)}$ at timelike
infinity. We show that in the case of the time-dependent background  Price's law holds only for
$\ell=0$ while for each $\ell\geq 1$ the tail decays as $t^{-(2\ell+2)}$.
\end{abstract}

\maketitle
The Einstein-massless scalar field system
\begin{equation}\label{einstein}
    G_{\alpha\beta}=8\pi \left(\nabla_{\alpha}\phi \nabla_{\beta}\phi-\frac{1}{2}g_{\alpha\beta}
    (\nabla_{\mu}\phi\nabla^{\mu}\phi)\right)\,,\qquad
    g^{\alpha\beta}\nabla_{\alpha}\nabla_{\beta} \phi=0\,,
\end{equation}
restricted to spherical symmetry, has been serving as an important theoretical laboratory for the
investigation of nonlinear gravitational phenomena in a rather simple $1+1$ dimensional setting.
For this system Christodoulou proved that a generic spherically symmetric solution  settles down
asymptotically either to Minkowski spacetime (for small data) \cite{chr1}, or to a Schwarzschild
black hole (for large data) \cite{chr2}. The first reliable numerical simulations of the
late-time asymptotics of this relaxation process have been done by Gundlach, Price, and Pullin
(GPP) \cite{gpp2}. They found that, regardless  of the endstate of evolution, the scalar field
develops a tail which falls off as $t^{-3}$ near timelike infinity (for compactly supported
initial data).

In a recent paper \cite{bcr5} we revisited this problem to emphasize that the asymptotic
convergence to a static equilibrium (Minkowski or Schwarzschild) is an essentially nonlinear
phenomenon which cannot, despite many assertions to the contrary in the literature, be properly
described by the theory of linearized perturbations on a fixed static asymptotically flat
background (Price's tails \cite{price, gpp1}). This is particularly evident for dispersive
solutions which asymptote Minkowski spacetime. In that case
  the quantitative characteristics of the tail
 (the decay rate and the amplitude) can be obtained using nonlinear perturbation expansion
 \cite{bcr5}. Since some details of this formal calculation will be needed below, let us now
 briefly summarize it.
In the  parametrization
\begin{equation}\label{metric}
ds^2 = \left(1-\frac{2m(t,r)}{r}\right)^{-1} \left(-e^{2\beta(t,r)} dt^2 + dr^2\right) + r^2
(d\vartheta^2+\sin^2\!{\vartheta}\, d\varphi^2)\,,
\end{equation}
the system (\ref{einstein}) takes a particularly convenient form (below an overdot denotes
$\partial/\partial t$ and a prime denotes $\partial/\partial r$)
\begin{eqnarray}
\label{h-constraint} m'&=& 2\pi\, r (r-2m) (\phi'^2 + e^{-2\beta} \dot\phi^2)\,,
\\
\dot{m}&=& 4\pi \,r (r-2m) \dot \phi \,\phi'\,,
\\
\label{s-condition} \beta' &=& \frac{2m}{r(r-2m)} \,,
\\
\left(e^{-\beta}\dot\phi\right)^{\cdot}&=&\frac{1}{r^2}\left(r^2 e^{\beta}
    \phi'\right)'\,.
\end{eqnarray}
Consider small and compactly supported initial data
 $(\phi,\dot\phi)_{t=0} = (\varepsilon f(r),
\varepsilon g(r))$. Then,
 up to the order $\mathcal{O}(\varepsilon^3)$, we have
\begin{equation}\label{phi1}
\phi =\varepsilon \phi_1+\varepsilon^3 \phi_3,\qquad m= \varepsilon^2 m_2,\qquad \beta=
\varepsilon^2 \beta_2,
\end{equation}
where $\phi_1$ satisfies the flat-space radial wave equation
\begin{equation}\label{eqphi1}
\Box \phi_1:=\ddot \phi_1-\phi_1''-\frac{2}{r} \phi_1'=0,\qquad (\phi_1,\dot
\phi_1)_{t=0}=(f(r),g(r)),
\end{equation}
while the second-order perturbations of the metric functions satisfy
\begin{equation}\label{eqmetric}
 m'_2 =
2\pi r^2 (\dot{\phi}_1^2 + \phi_1'^2),\qquad \beta'_2 = \frac{2 m_2}{r^2}\,.
\end{equation}
Solving equation (\ref{eqphi1}) and then (\ref{eqmetric}) we get for $t>R$ (where $R$ is the
radius of support of initial data)
\begin{eqnarray}\label{m2}
\phi_1(t,r)&=&\frac{a(t-r)}{r},\\
m_2(t,r) &=& 2\pi \left( 2 \int \limits_{t-r}^\infty a'^2(s) \, ds - \frac {a^2(t-r)}
{r}\right)\,,\\
\beta_2(t,r) &=& 4\pi \left( -\frac{2}{r} \int \limits_{t-r}^{\infty} a'^2(s) \, ds + 2 \int
\limits_{t-r}^{\infty} \frac {a'^2(s)} {t-s} \, ds - \int \limits_{t-r}^{\infty} \frac {a^2(s)}
{(t-s)^3} \, ds \right) \,,
\end{eqnarray}
where the initial-data-generating function $a(u)$ vanishes for $|u|>R$. The third-order
perturbation of the scalar field $\phi_3$ satisfies the inhomogeneous wave equation (with zero
initial data)
\begin{equation}
\label{phi3} \Box \phi_3  = 2 \beta_2 \ddot{\phi}_1 + \dot{\beta}_2 \dot{\phi}_1+ \beta'_2
\phi'_1 =:S(t,r)\,.
\end{equation}
The source $S(t,r)$ is already known from (10-12) so we can use the Duhamel formula
\begin{equation}
\label{duh0} \phi_3(t,r)= \frac {1} {2 r} \int \limits_{0}^{t} d\tau \int
\limits_{|t-r-\tau|}^{t+r-\tau} \rho\, S(\tau,\rho) \,d\rho\,,
\end{equation}
to obtain the  asymptotic behavior for large retarded times
\begin{equation}\label{gamma0}
\phi(t,r)\simeq \varepsilon^3 \phi_3(t,r) \sim \frac {\varepsilon^3 \Gamma_0\,
t}{(t^2-r^2)^2}\,,\qquad
 \Gamma_0 = -2^5\pi \int \limits_{-\infty}^{+\infty} a(u) \int \limits_{u}^{+\infty}
(a'(s))^2 ds \,du\,.
\end{equation}
 We refer the reader to \cite {bcr5} for more details about this calculation and numerical
 evidence.

After this introduction, we are ready to discuss an interesting model for investigating linear
\emph{nonspherical} tails on a fixed dynamical background. This model, proposed by GPP
\cite{gpp2}, involves a nonspherical test massless scalar field $\psi$ which propagates on the
spacetime (\ref{metric}) generated by the self-gravitating field $\phi$. Since the dynamics of
$\psi$ is linear
 and the background is spherically symmetric, one may decompose $\psi$ into spherical harmonics
\begin{equation}\label{decomp}
    \psi(t,r,\vartheta,\varphi)=\sum_{\ell\geq 0, |m|\leq \ell}
     \psi_{\ell m}(t,r) \,Y^m_{\ell}(\vartheta,\varphi)\,,
\end{equation}
and analyze the evolution of each multipole separately
\begin{equation}\label{eqpsi}
    \left(e^{-\beta}\dot \psi_{\ell m} \right)^{\cdot}-\frac{1}{r^2}\left(r^2 e^{\beta}
    \psi_{\ell m}'\right)'+e^{\beta}\frac{\ell(\ell+1)}{r(r-2m)}\,
    \psi_{\ell m}=0\,.
\end{equation}
GPP conjectured\footnote{GPP considered the characteristic initial value problem while we are
studying the Cauchy problem, however this difference do not affect the asymptotics of tails.}
that for compactly supported initial data the multipoles have the tail $\psi_{\ell m}(t,r) \sim
t^{-(2\ell+3)}$ at timelike infinity ($t\rightarrow \infty$ at a fixed $r$), in accord with Price's
law on a fixed Schwarzschild background, even though the actual spherical background is time
dependent and its Bondi mass decreases (to a positive value in the collapsing case or to zero in
the dispersive case). It seems that GPP's conjecture was based more on belief than numerical
evidence, because for the first few multipoles the following power-law exponents of the tail were
reported numerically (see Fig.12 in \cite{gpp2}): $-2.77$ ($\ell=0$), $-3.95$ ($\ell=1$), $-5.94$
($\ell=2$), and $-8.34$ ($\ell=3$).

The purpose of this note is to point out that in this model (and for other time-dependent
backgrounds) Price's law ({\em i.e.}, $t^{-(2\ell+3)}$ decay) holds only for $\ell=0$, while for
$\ell \geq 1$ the power-law exponent of the tail is equal to $-(2\ell+2)$ (as was clearly indicated
by GPP's own numerics). To show that we shall compute the late-time asymptotic behavior of
$\psi_{\ell m}(t,r)$ (for smooth initial data compactly supported in a ball of radius $R'$) along
similar lines as described above for the field $\phi$. The perturbation expansion has the form
$\psi_{\ell m}=\psi_0+\varepsilon^2 \psi_2+\dots$, where for convenience of notation we dropped the
multipole indices  on iterates. At the zero order we have
\begin{equation}
\label{psi1} \Box_{(\ell)} \psi_0 := \ddot \psi_{0}-\psi_{0}''-\frac{2}{r}
\psi_{0}'+\frac{\ell(\ell+1)}{r^2}\psi_{0}=0\,,\qquad (\psi_0,\dot \psi_0)_{t=0}=(\psi_ {\ell
m},\dot \psi_{\ell m})_{t=0}\,,
\end{equation}
which for $t>R'$ is solved by
\begin{equation}
\label{f1} \psi_0 (t,r) =
 \frac{1}{r}\,\sum_{k=0}^{l}  \frac {(2\ell-k)!}
{k!(\ell-k)!} \frac {b^{(k)}(t-r)}{(2r)^{\ell-k}}\,,
\end{equation}
where the initial-data-generating function $b(u)$ vanishes for $|u|>R'$  (the superscript in
round brackets denotes the $k$-th derivative). At the second order we get
\begin{equation}
\label{psi2} \Box_{(\ell)} \psi_2  = 2 \beta_2 \ddot{\psi}_0 + \dot{\beta}_2 \dot{\psi}_0+ \beta'_2
\psi'_0 - \frac{2\ell(\ell+1) m_2}{r^3}\psi_0 =:S_{\ell}(t,r)\,,\qquad (\psi_2,\dot
\psi_2)_{t=0}=(0,0)\,.
\end{equation}
Substituting (11), (12), and (\ref{f1}) into the Duhamel formula (where $P_{\ell}(x)$ is the
Legendre polynomial of degree $\ell$)
\begin{equation}
\label{duhL} \psi_2(t,r)= \frac {1} {2 r} \int \limits_{0}^{t} d\tau \int
\limits_{|t-r-\tau|}^{t+r-\tau} \rho\, P_{\ell}\left(
\frac{r^2+\rho^2-(t-\tau)^2}{2r\rho}\right)\,S_{\ell}(\tau,\rho) \,d\rho\,,
\end{equation}
 we get (for large retarded times) for $\ell=0$:
 \begin{equation}\label{tailpsi0}
 \psi_{00}(t,r) \simeq \varepsilon^2 \psi_2(t,r) \sim  \frac {\varepsilon^2 B_0 \,t} {(t^2-r^2)^2}\,,\qquad
    B_0 = -2^5\pi \int \limits_{-\infty}^{+\infty} b(u) \int \limits_{u}^{+\infty} (a'(s))^2 ds
\,du\,,
\end{equation}
and for $\ell\geq 1$ (see \cite {bcrz} for technical details of calculation in the $\ell\geq 1$ case):
\begin{equation}\label{tailpsiL}
 \psi_{\ell m}(t,r) \simeq \varepsilon^2 \psi_2(t,r) \sim  \frac {\varepsilon^2 B_{\ell}\, r^{\ell}} {(t^2-r^2)^{\ell+1}}\,,\qquad
  B_{\ell}=(-1)^{\ell}\frac{2^{\ell+3} \ell! \pi}{2\ell+1} \int \limits_{-\infty}^{+\infty}
  \left( \frac{\ell^2(\ell - 1)}{2\ell-1} a^2(u)
   b''(u) - 2 \ell^2 (a'(u))^2 b(u) \right) \,du\,.
 \end{equation}
\begin{figure}[ht]
\begin{tabular}{cc}
\includegraphics[width=0.45\textwidth]{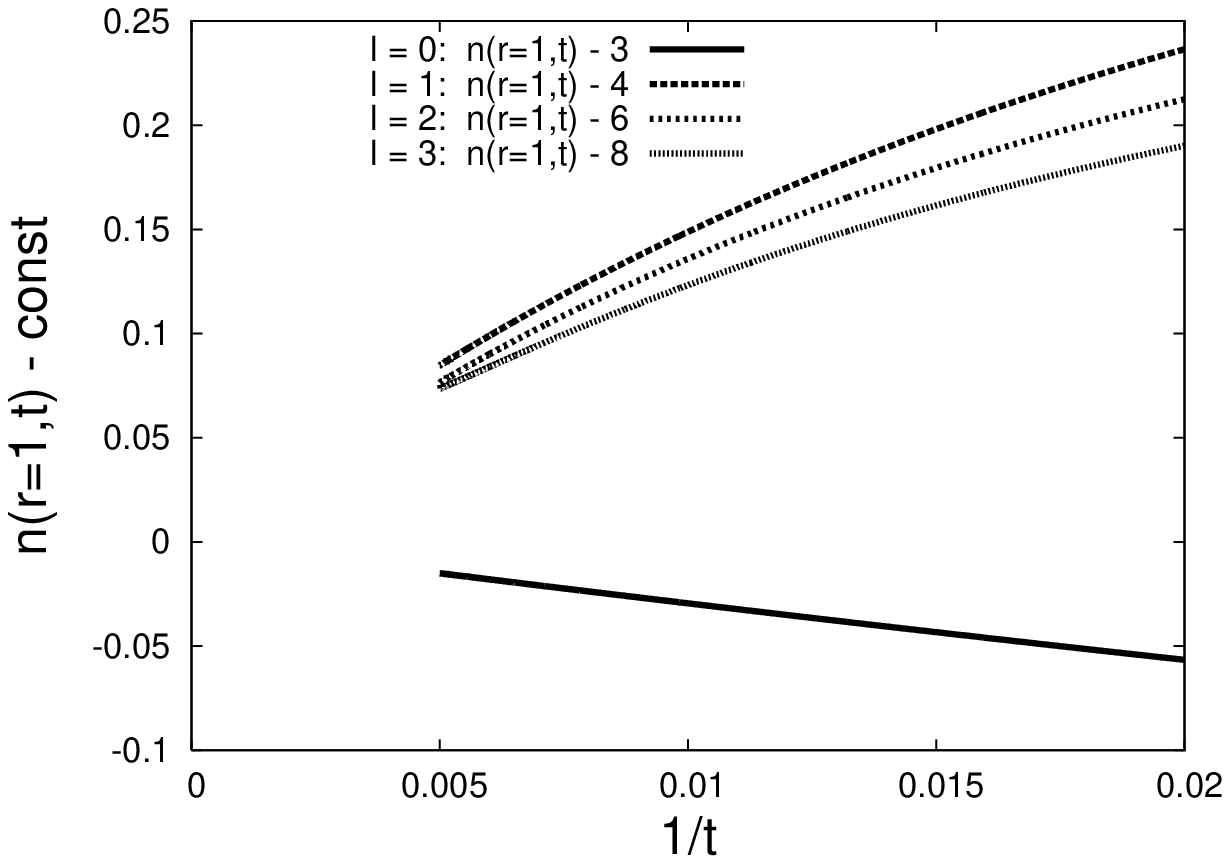}
&
\includegraphics[width=0.45\textwidth]{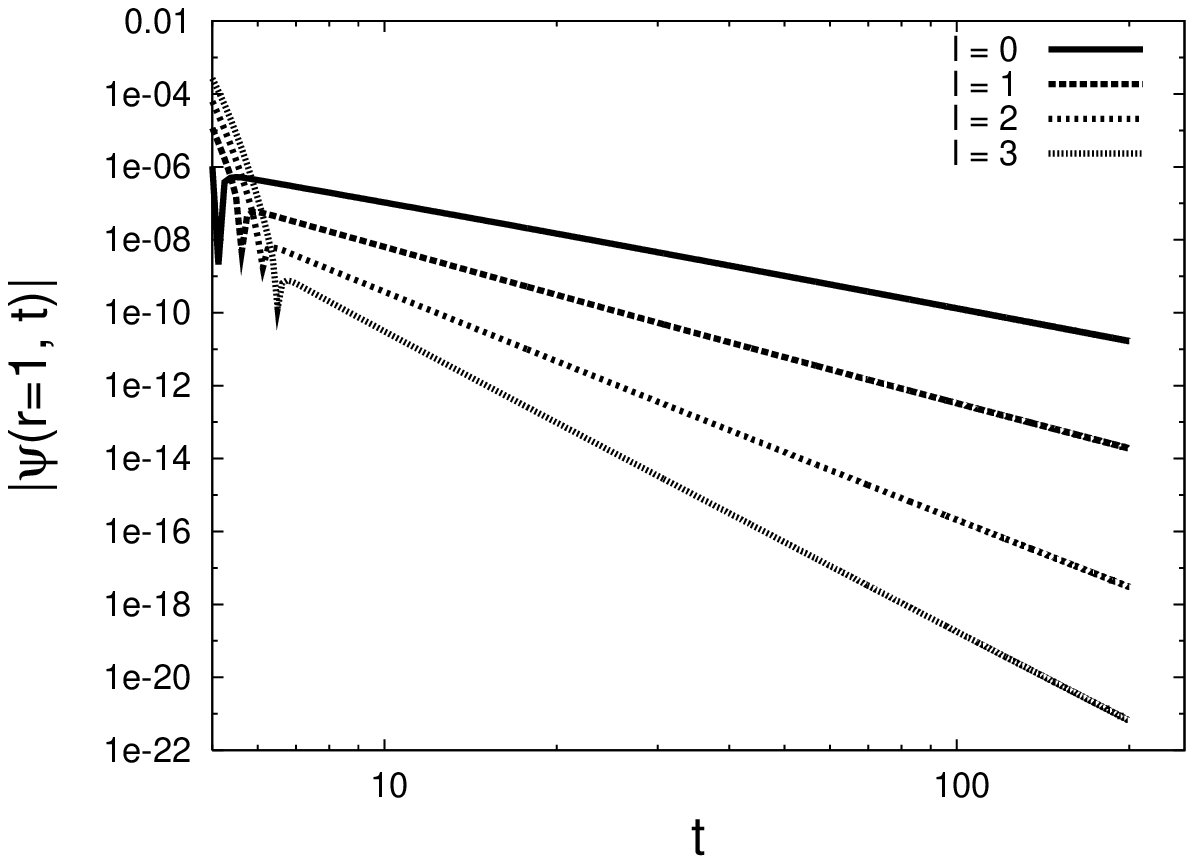}
\\
\end{tabular}
\caption{ Left panel: The difference between the local power index and the theoretical prediction:
$n(r=1,t)-3$ for $\ell=0$ and $n(r=1,t)-2(\ell+2)$ for $\ell=1,2,3\,$, as a function of $1/t$.
Right panel: The log-log plots of $|\psi(t,\,r=1)|$ for $\ell=0,1,2,3$.
 Both panels correspond to initial data generated
 by $a(x)=\exp(-x^2)/\sqrt{2\pi} $
 (for $\phi_1$) and $b(x)=x^2 \exp(-x^2)$ (for $\psi_0$), with $\varepsilon = 2^{-8}$.}
\end{figure}
\begin{table}[ht]
\centering
\begin{tabular}{|c||c|c|c||c|c||c|}
\hline
$\varepsilon$ & \multicolumn{3}{c||}{Numerics: LPI data} & \multicolumn{2}{c||}{Theory (third order)}
& Numerics: $\psi$ data \\
\cline{2-7}
 & $c$ & $d$ & $\gamma$ & $\gamma$  & $B$  & $B$ \\
\hline
\hline
\multicolumn{7}{|c|}{$\ell=0$}\\
\hline
\hline
$2^{-12}$ & -3.068 & 6.056 & $3.000$ & $3$ & $-5.296 \times 10^{-4}$ & $-5.293 \times 10^{-7}$ \\
\hline
$2^{-10}$ & -3.064 & 5.877 & $3.000$ & $3$ & $-8.474 \times 10^{-4}$ & $-8.468 \times 10^{-6}$ \\
\hline
$2^{-8}$ & -3.064 & 5.867 & $3.000$ & $3$ & $-1.356 \times 10^{-4}$ & $-1.355 \times 10^{-4}$ \\
\hline
\hline
\multicolumn{7}{|c|}{$\ell=1$}\\
\hline
\hline
$\,\,2^{-12}\,\,$ & $\,\,16.73\,\,$ & $\,\,-134.7\,\,$ & $\,\,4.008\,\,$ & $\,\,\,\,4\,\,\,\,$ & $1.084 \times 10^{-7}$ &  $1.145 \times 10^{-7}$ \\
\hline
$2^{-10}$ & 16.73 & -134.7 & 4.008 & 4 & $1.735 \times 10^{-6}$ & $1.833 \times 10^{-6}$ \\
\hline
$2^{-8}$ & 16.73 & -134.7 & 4.008 & 4 & $2.776 \times 10^{-5}$ & $2.932 \times 10^{-5}$ \\
\hline
\hline
\multicolumn{7}{|c|}{$\ell=2$}\\
\hline
\hline
$\,\,2^{-12}\,\,$ & $\,\,15.87\,\,$ & $\,\,-139.4\,\,$ & $\,\,6.005\,\,$ & $\,\,\,\,6\,\,\,\,$ & $-6.940 \times 10^{-7}$ & $-7.193 \times 10^{-7}$ \\
\hline
$2^{-10}$ & 15.87 & -139.4 & 6.005 & 6 & $-1.110 \times 10^{-5}$ & $-1.151 \times 10^{-5}$ \\
\hline
$2^{-8}$ & 15.87 & -139.4 & 6.005 & 6 & $-1.777 \times 10^{-4}$ & $-1.841 \times 10^{-4}$ \\
\hline
\hline
\multicolumn{7}{|c|}{$\ell=3$}\\
\hline
\hline
$\,\,2^{-12}\,\,$ & $\,\,13.28\,\,$ & $\,\,-110.7\,\,$ & $\,\,8.012\,\,$ & $\,\,\,\,8\,\,\,\,$ & $6.023 \times 10^{-6}$ & $6.525 \times 10^{-6}$ \\
\hline
$2^{-10}$ & 13.29 & -111.0 &  8.012 & 8 & $9.637 \times 10^{-5}$ & $1.044 \times 10^{-4}$ \\
\hline
$2^{-8}$ & 13.31 & -111.4 & 8.012 & 8 & $1.542 \times 10^{-3}$ & $1.669 \times 10^{-3}$ \\
\hline
\hline
\end{tabular}
\caption{The comparison of analytic and numerical decay rates and amplitudes of the tails at
timelike infinity for $\ell=0,1,2,3$, for initial data generated by $a(x)=\exp(-x^2)/\sqrt{2\pi}$
(for $\phi_1$) and $b(x)=x^2 \exp(-x^2)$ (for $\psi_0$), at $r=1$. The theoretical prediction is
$B=\varepsilon^2 B_{\ell}$, with $B_{0}$ given in (\ref{tailpsi0}) and $B_{\ell}$ given in
(\ref{tailpsiL}) for $\ell \geq 0$. Fits were made on the interval $50 \leq t \leq 200$.}
\end{table}
The numerical verification of these formulae is summarized in Figure~1 and Table~1.

\noindent We fit our numerical data with the formula
\begin{equation}
\label{numpsi} \psi (t,r) = B t^{-\gamma} \exp \left(c/t + d/t^2 \right)\,,
\end{equation}
which gives the local power index (LPI) \cite{bo}
\begin{equation}
\label{lpi} n(t,r) := -t \dot \psi(t,r) / \psi(t,r) = \gamma + c/t + 2d/t^2\,.
\end{equation}
Our fitting procedure proceeds in two steps. First, from the local power index data on the interval
$1/t \leq 1/50$ (the left panel of Figure~1) we fit $\gamma$, $c$ and $d$ in (\ref{lpi}). Next,
having determined $\gamma$, $c$ and $d$ in this way, we fit $B$ in (\ref{numpsi}) from $\psi$ data
on the interval $50 \leq t$ (the right panel of Figure~1).
 The results of this procedure are given in Table~1.

  It is instructive to compare the tail (\ref{tailpsiL})
with the tail of a massless scalar field propagating on a fixed  asymptotically flat {\em static}
background. The latter can be readily obtained from the Duhamel formula (\ref{duhL}) applied to
the source (\ref{psi2}) with $m_2=M$ and $\beta_2=-2M/r$, where $M$ is the total mass. The result
(valid for all $\ell$) reads
 \begin{equation}\label{price}
  \psi_{\ell m}(t,r) \sim  \frac {C_{\ell}\, r^{\ell} t} {(t^2-r^2)^{\ell+2}}\,,\qquad
  C_{\ell}=-M 2^{\ell+3} (\ell+1)!  \int \limits_{-\infty}^{+\infty} b(u) du\,.
\end{equation}
 This is the celebrated Price tail \cite{price} (as far as we know, first obtained in the form (\ref{price})
 by Poisson \cite{poisson}). It is worth stressing that the formula (\ref{price}) yields a good
  approximation
 for the amplitude of the tail provided that both an observer and initial data lie in a weak
 field region where $M/r$ is small. Note that for $\ell=0$ the formula (\ref{tailpsi0})
 takes the form (\ref{price})
 when the total mass $M$ is replaced by the weighted average over the Bondi mass
 $M(u)=4\pi \int \limits_{u}^{+\infty} (a'(s))^2
 ds$. For $\ell\geq 1$ Price's tail decays by one power faster than that in (\ref{tailpsiL}) which, on a
 technical level, is due an extra cancelation in the integration by parts of Duhamel's
 formula.

 \vskip 0.2cm \noindent \textbf{Acknowledgments:}  We acknowledge support by the MNII
grants: NN202 079235 and 189/6.PRUE/2007/7.

\end{document}